\documentclass[nonacm,sigconf]{acmart}
\usepackage{graphicx}

\acmISBN{978-1-4503-XXXX-X/18/06}

\begin{document}
\title{Uses of Active and Passive Learning in Stateful Fuzzing}


\author{Cristian Daniele}
\email{cristian.daniele@ru.nl}
\affiliation{%
  \institution{Radboud University}
  \city{Nijmegen}
  \country{The Netherlands}}

\author{Seyed Behnam Andarzian}
\email{seyedbehnam.andarzian@ru.nl}
\affiliation{\institution{Radboud University}
\city{Nijmegen}
\country{The Netherlands}
}

\author{Erik Poll}
\email{erikpoll@cs.ru.nl}
\affiliation{%
  \institution{Radboud University}
  \city{Nijmegen}
  \country{The Netherlands}}


\begin{abstract}
  This paper explores the use of active and passive learning, i.e.\ active and passive techniques to infer state machine models of systems, for fuzzing. Fuzzing has become a very popular and successful technique to improve the robustness of software over the past decade, but stateful systems are still difficult to fuzz. Passive and active techniques can help in a variety of ways: to compare and benchmark different fuzzers, to discover differences between various implementations of the same protocol, and to improve fuzzers.


  \end{abstract}

\maketitle

\section{Introduction}\label{sec:intro}

Fuzzing (or fuzz testing) is a testing technique used in the late eighties to find vulnerabilities in UNIX utilities by sending malformed messages. Despite the technique being known for more than 30 years, only recently the software security community showed interest in stateful fuzzers, i.e.\ fuzzers specifically tailored to stateful systems.

Research into methods to infer state models refers back to the 1980s as well, notably with the L* algorithm \cite{angluin1987learning} as a black-box technique to infer the state model of a system by so-called active learning, i.e.\ interacting with the system and observing its responses.

Passive learning is another approach to infer state models of systems. It does not require interaction with the system but just needs a set of collected traces~\cite{yang2019improving}.

These three techniques (stateful fuzzing, active learning and passive
learning) can be combined in various ways to improve the security testing
of stateful systems. Having a state model of the System Under Test (SUT)
can be an important advantage for stateful fuzzing, and both active
and passive learning can be used to supply such a model (see
Section~\ref{sec:effectiveness}).  Moreover, active and passive learning can be
used to benchmark stateful fuzzers (see
Section~\ref{sec:benchmarking}) or for differential testing (see
Section~\ref{sec:differential}).

\subsection{Fuzzing and Stateful Fuzzing}


Fuzzing is a testing technique used to find vulnerabilities in software~\cite{zhu2022fuzzing,li2018fuzzing}. In the last decade, especially since the advent of AFL~\cite{afl}, the use of fuzzers have become very successful, revealing many security flaws.

To fuzz a SUT we feed it randomly generated, often
malformed inputs to check if these trigger bugs. 
Stateful fuzzing -- i.e.\ fuzzing a stateful system -- is more
challenging: instead of just sending a message we may need to send a
sequence of messages (what we will call a trace) to get the SUT in the
right state where a bug can be triggered~\cite{daniele2024fuzzers}. Knowing the state model of
the SUT can help as we can make sure we visit all the states,
effectively fuzzing the SUT in each state. 

\subsection{Active Learning}
Active learning infers the state model of a SUT by interacting with
it~\cite{tharwat2023survey}. The idea is that we gradually improves our understanding of the
model by trying out sequences of inputs and checking if the observed
output correspond with the expected behaviour, and if not, refining
the state machine model for the behaviour.

Active learning can be very accurate but also very slow, as it
involves exhaustively trying our all the possible traces up to a given
length.  The state space can explode because of the number of possible
messages and the size of the state models. It can be regarded as a
limited form of fuzzing~\cite{de2015protocol}, where we only mutate
the order of a fixed set of messages but not these messages
themselves. But a core difference between active learning and stateful
fuzzing is that the former aims to infer the state model of the SUT,
while the latter aims to trigger bugs.

Active learning has been proven extremely effective and broadly applicable thanks to its black-box nature.  

\subsection{Passive Learning}\label{sec:passive-and-fuzzing}

Passive learners are quite different from the stateful fuzzers and the
active learner tools.  While fuzzers and active learning tools
\textit{actively} send messages to the SUT, passive learners take a
set of traces that have been collected beforehand as
input to then infer a state machine~\cite{yang2019improving}. During this process, also known as
\textit{grammatical inference}, there is no further interaction with the SUT.
The set of traces used could be produced by using a stateful fuzzer.

Passive learning used in combination with a stateful fuzzer can
gives a good approximation of the state model much faster than active
learning. It could even produce a more detailed model than an active
learner tool since the traces given to the passive learner may also
contain malformed messages (and not just malformed traces of correctly
formed messages). However, that depends heavily on the effectiveness
of the (stateful) fuzzer.

\section{Combinations} \label{sec:combinations}

Stateful fuzzing, active learning and passive learning can be combined
to achieve different goals.

\subsection{Improving the effectiveness of stateful fuzzers}
\label{sec:effectiveness}

Some stateful fuzzers already use passive or active learning (\cite{pham2020aflnet,bastani2017synthesizing,aichernig2021learning}), but most of them do not. 
Since active and passive learning techniques work in a black-box
fashion, they can be easily added to existing stateful fuzzers in order to improve their state-awareness.

\subsection{Benchmarking of stateful fuzzers}
\label{sec:benchmarking}

Benchmarking fuzzers is challenging and benchmarking stateful fuzzers
is even more so \cite{hazimeh2020magma}.  Knowing the state model can
help here as it allows us to count how many states are visited during
the fuzzing and how heavily each state has been fuzzed.  For example,
if \textit{fuzzer A} reaches 3 of the 8 states and \textit{fuzzer B} 7
of the 8 states we can argue that \textit{fuzzer B} is more effective
than \textit{fuzzer A} in exploring the state model.  Both active and
passive learning can help in providing the state model than we can
then use to measure the coverage.

\begin{figure}[t]
  \begin{minipage}{0.5\textwidth}
      \centering
      \includegraphics[width=0.9\linewidth]{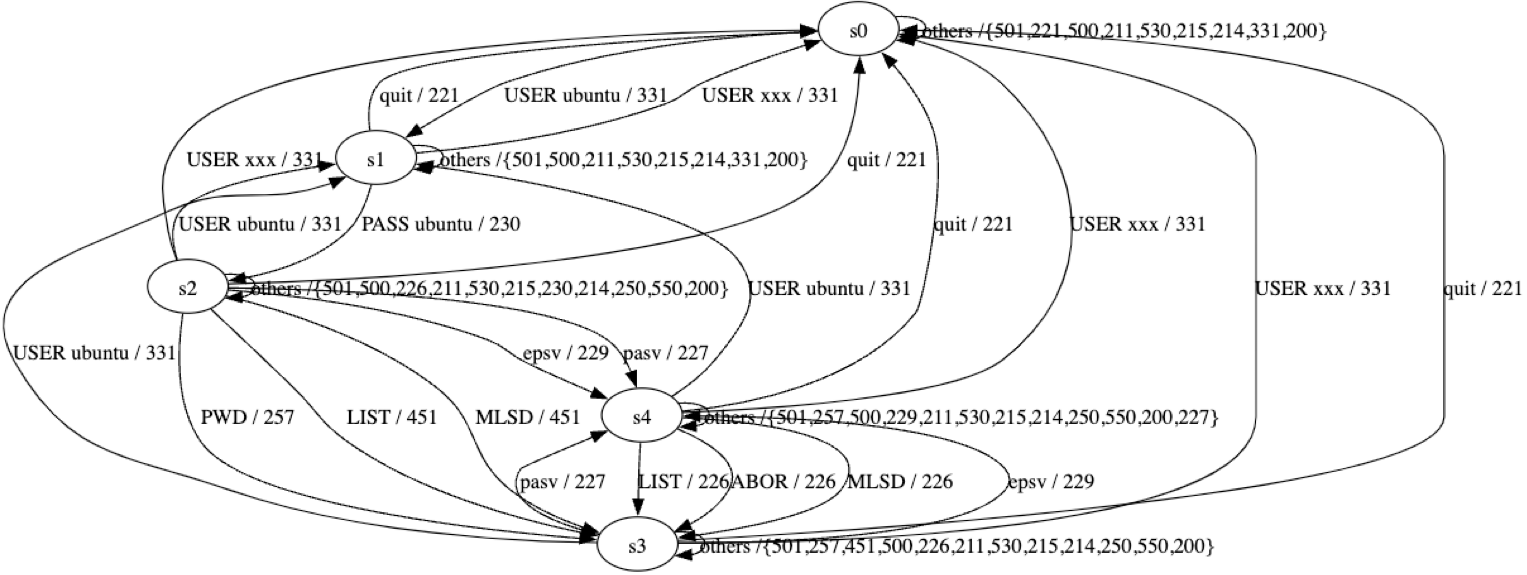}
      \caption{State model of LightFTP inferred by the active learning
      tool LearnLib\cite{LearnLib}}
      \label{fig:LightFTP-LearnLib}
  \end{minipage}
  \hfill
  \begin{minipage}{0.45\textwidth}
      \centering
      \includegraphics[width=0.8\linewidth]{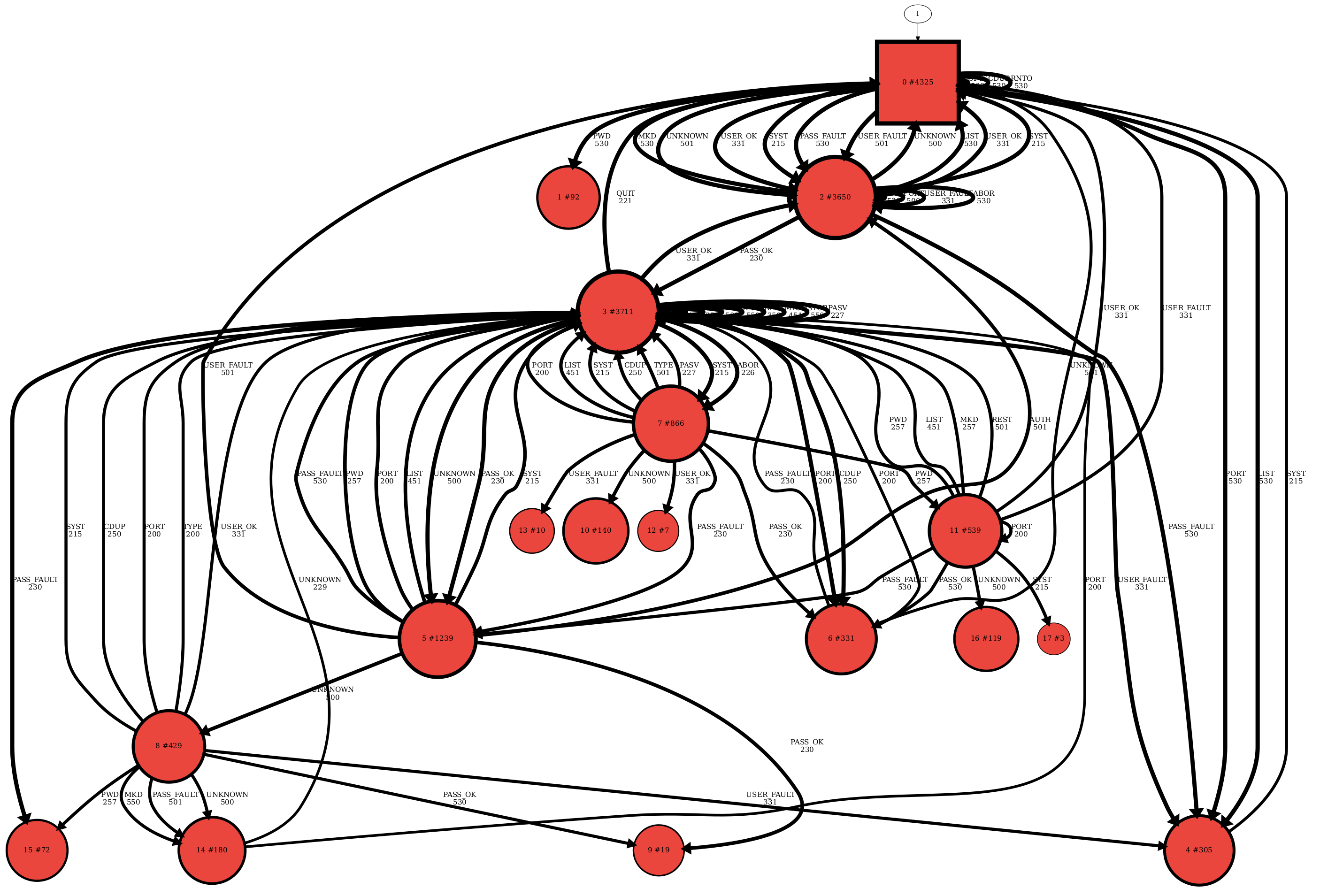}
      \caption{State model of LightFTP inferred by AFLNet
      and the passive learning tool FlexFringe\cite{verwer2017flexfringe}}
      \label{fig:FlexFringe}
  \end{minipage}
\end{figure}

The state models inferred by active and passive learning tools can be
different (e.g.\ see 
Fig.~\ref{fig:LightFTP-LearnLib} and \ref{fig:FlexFringe}). As
already mentioned in Section~\ref{sec:passive-and-fuzzing}, the set of
traces given as input to a passive learner may not only involve
mutations in the order of the messages (which active learning would
also explore) but may also involve mutations of individual messages
(which active learners typically do not explore, but which most
fuzzers will).  Differences between state models inferred by active
and passive learning may point to flaws in the SUT, more specifically
in the program logic when it comes to handling malformed messages.

\subsection{Differential testing}\label{sec:differential}

\begin{figure}[t]
  \centering
  \begin{minipage}{0.5\textwidth}
      \centering
      \includegraphics[width=\linewidth]{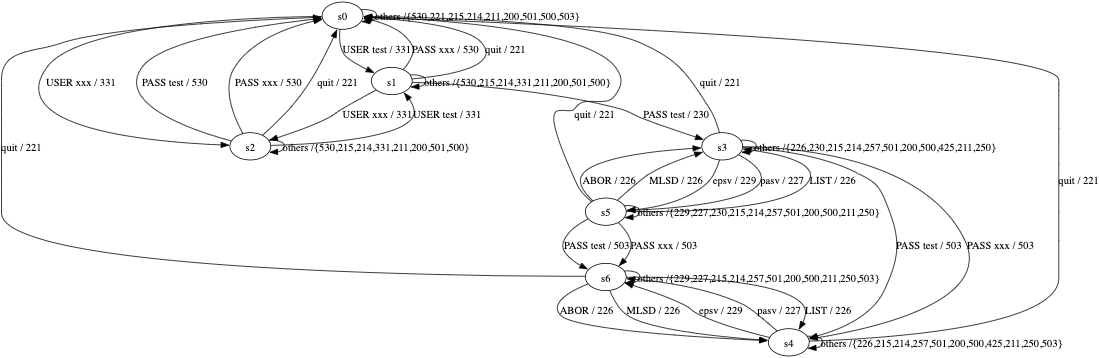}
      \caption{State model for ProFTPd inferred by LearnLib}
      \label{fig:ProFTPd-LearnLib}
  \end{minipage}
  \hfill
  \begin{minipage}{0.5\textwidth}
      \centering
      \includegraphics[width=0.7\linewidth]{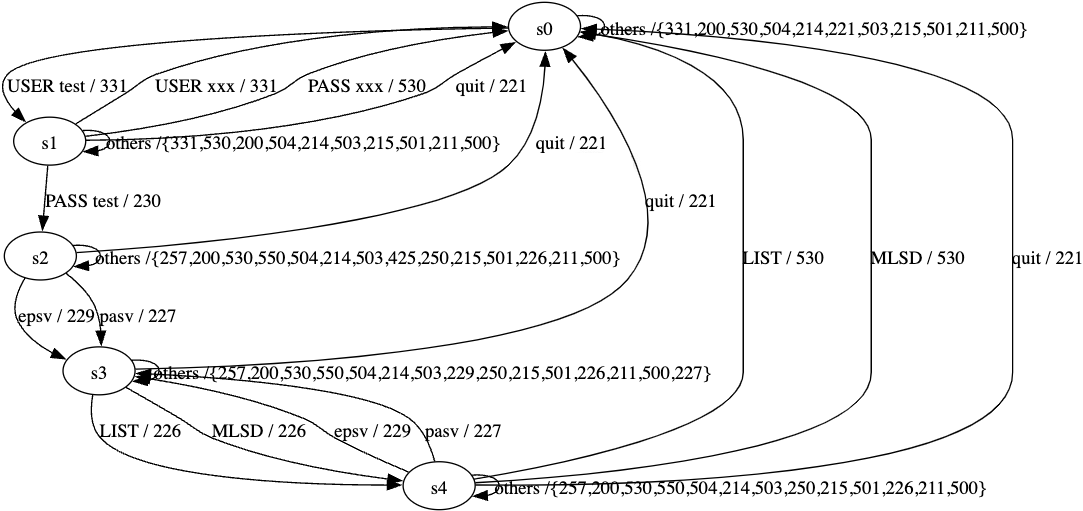}
      \caption{State model for PureFTPd inferred by LearnLib}
      \label{fig:PureFTPd-LearnLib}
  \end{minipage}
  \hfill
  \begin{minipage}{0.5\textwidth}
      \centering
      \includegraphics[width=\linewidth]{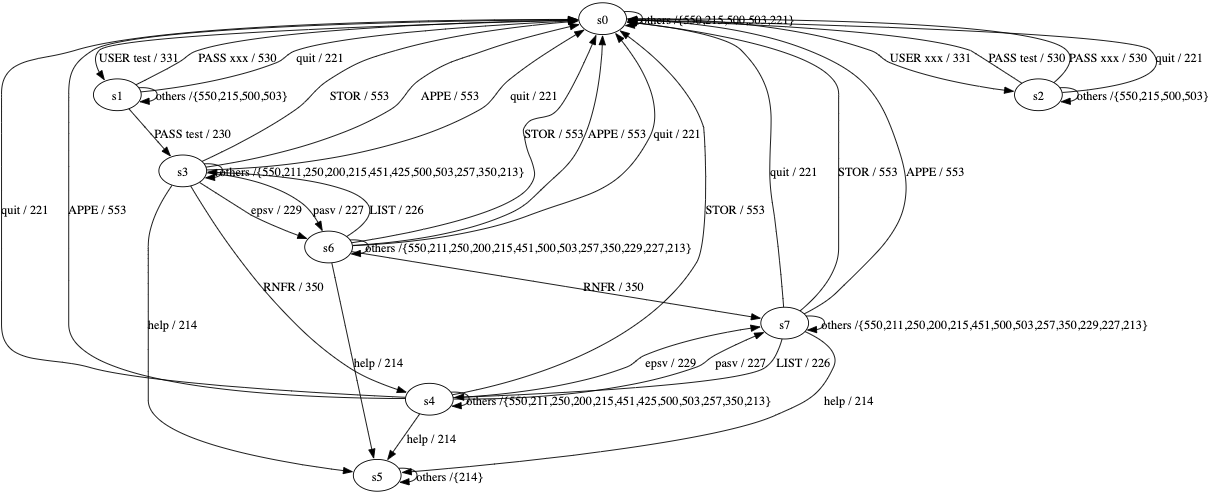}
      \caption{State model for bftpd inferred by LearnLib}
      \label{fig:bftpd-LearnLib}
  \end{minipage}

\end{figure}

Comparing the state models of different implementations of the same
protocol can also be valuable~\cite{mckeeman1998differential}.  As shown in
Fig.~\ref{fig:LightFTP-LearnLib},~\ref{fig:ProFTPd-LearnLib},~\ref{fig:PureFTPd-LearnLib}
and~\ref{fig:bftpd-LearnLib}, different implementations of FTP have
different state models.  Such differences between these state models
may point to flaws in the program logic or ambiguities in the
specification.

\section{Conclusions}\label{sec:conclusions}
This paper sheds light on the differences and similarities between
active learning, passive learning and stateful fuzzing and ways these
could be combined.  In particular, we suggest how stateful
fuzzers might benefit from models obtained by active or passive
learning to improve their effectiveness, allowing benchmarking
and do differential fuzzing.

\begin{acks}
  This research is funded by NWO as part of the INTERSECT project
  (NWA.1160.18.301).
\end{acks}

\bibliographystyle{ACM-Reference-Format}
\bibliography{bib}

\end{document}